\documentclass[final,5p,times,twocolumn]{elsarticle}

\usepackage{amssymb}
\usepackage{amsmath}
\usepackage{amsfonts}
\usepackage{graphicx}
\usepackage{empheq}
\usepackage{orcidlink}
\usepackage{enumitem}
\usepackage{xcolor}
\usepackage{hyperref}

\newcommand{\red}[1]{{#1}}
\definecolor{linkcolour}{HTML}{000066}
\hypersetup{colorlinks,breaklinks,
	urlcolor=linkcolour, 
	linkcolor=linkcolour,
	citecolor=linkcolour}

\journal{Optics Communications}

\begin{document}
	
	\begin{frontmatter}
		
		\title{Anisotropy in Fourier space optical memory effect correlations}
		
		\author[eee]{Niall Byrnes\orcidlink{0000-0002-1554-3820}}
		
		\author[eee,idmxs]{Matthew R. Foreman\corref{cor1}\orcidlink{0000-0001-5864-9636}}
		\ead{matthew.foreman@ntu.edu.sg}
		
		\cortext[cor1]{Corresponding author}
		
		\affiliation[eee]{organization={School of Electrical and Electronic Engineering, Nanyang Technological University},
			addressline={50 Nanyang Avenue}, 
			postcode={639798}, 
			country={Singapore}}
		
		\affiliation[idmxs]{organization={Institute for Digital Molecular Analytics and Science},
			addressline={59 Nanyang Drive}, 
			postcode={636921}, 
			country={Singapore}}
		
		\begin{abstract}
			We investigate anisotropy in Fourier-domain speckle correlations associated with the optical memory effect in disordered scattering media. Within a single scattering framework, we show that while the conventional memory effect constrains transverse wavevector shifts, the correlation strength also depends non-trivially on differences in the axial wavevector components. Our theory is supported by numerical simulations of a three-dimensional, single scattering medium, which show excellent agreement with theory. We extend the analysis to pseudo-correlations, demonstrating that analogous anisotropic behavior arises in the conjugate memory effect. Our results highlight the often overlooked role of axial disorder in scattered field correlations.
		\end{abstract}
		
		
	\end{frontmatter}

\section{Introduction}

The optical memory effect describes a fundamental correlation exhibited by speckle patterns produced by waves propagating through complex scattering media under small angular tilts of the incident field~\cite{PhysRevLett.61.834}. Since its initial discovery, substantial effort has been devoted to understanding its physical origin~\cite{Liu:19}, as well as its connections to other key aspects of wave propagation in disordered media, including time-reversal symmetry \cite{PhysRevB.41.2635}, polarization~\cite{Berkovits_1990}, and scattering eigenchannels~\cite{PhysRevLett.123.203901}. \red{The memory effect is now a central tool in speckle correlation imaging~\cite{Katz2014}, and remains the subject of active research aimed at extending its practical applications~\cite{10.1117/1.AP.8.3.036009}.} A wider range of memory effects have also been established, demonstrating that speckle correlations persist under more general angular tilts~\cite{PhysRevX.11.031010}, combinations of shifts and tilts of the incident field~\cite{Osnabrugge:17}, and other types of transformations, such as frequency shifts~\cite{Zhang:21}, axial displacements~\cite{Zhu:20}, rotations~\cite{Amitonova:15}\red{, and radial shifts in optical fibers~\cite{10.1063/5.0303181}.}

A scattered field emerging from a complex medium is typically expressed as a sum of many partial fields, each associated with a distinct sequence of scattering events within the medium. The statistical properties of the scattered fields are obtained by averaging appropriate combinations of these partial fields over different physical realizations of the underlying disorder. The memory effect in particular arises as a constructive interference phenomenon  when averaging over the transverse spatial dimensions of the scattering medium~\cite{Byrnes02012026}. \red{The additional dependence arising from averaging over the axial (longitudinal) disorder, on the other hand, is often neglected. This dependence gives rise to anisotropic correlations in Fourier space, even when the transverse memory effect condition is satisfied. Although this effect was reported in Ref.~\cite{Byrnes02012026}, its geometric
structure and consequences for wide angle correlation measurements
remain largely unexplored. In this work, we characterize the geometry of these anisotropic correlations, identify the curves along which correlations are strongest, and examine their dependence on the illumination and observation geometry. We support our theoretical analysis with numerical simulations.}

\section{Theory}

We consider a three-dimensional slab centered at the origin with spatial extents $L_x, L_y,$ and $L_z$ along the Cartesian axes. We define the $x$ and $y$ axes as transverse directions, the $z$ axis as the axial direction, and assume that $L_x,L_y \gg L_z$. The interior of the slab is populated with $N$ randomly distributed point scatterers. The slab is illuminated by a monochromatic, unit-amplitude plane wave $E_i(\mathbf{r}) = \exp(i\mathbf{k}_i\cdot \mathbf{r})$, with incident wavevector $\mathbf{k}_i$. Note that while we neglect polarization for simplicity, our final results remain valid in a full vectorial treatment~\cite{ByrnesPhD, ByrnesCPC}. In an arbitrary measurement plane $z=z_m$ on either side of the slab, the scattered field can be decomposed into a spectrum of plane waves~\cite{Mandel_Wolf_1995}. Assuming that $L_z$ is smaller than the medium's scattering mean free path, the plane wave component of the scattered field propagating with wavevector $\mathbf{k}_j$ has complex amplitude given by~\cite{Byrnes02012026}
\begin{align}\label{eq:scattered-field-angular-spectrum}
	\widetilde{E}_{ji}\sim \sum_{n=1}^N f_n(\mathbf{k}_j, \mathbf{k}_i)\exp(i\mathbf{r}_n\cdot[\mathbf{k}_i - \mathbf{k}_j]),
\end{align}
where $n$ enumerates the scatterers, $\mathbf{r}_n$ is the location of the $n$'th scatterer, and $f_n$ is a form factor for the $n$'th scatterer.

The optical memory effect manifests in the field \red{covariance $C_{ijuv} = \langle \widetilde{E}_{ji}\widetilde{E}_{vu}^* \rangle - \langle \widetilde{E}_{ji}\rangle \langle \widetilde{E}_{vu}^* \rangle$}, where $u$ and $v$ denote an additional pair of incident and scattered plane wave directions, and \red{$\langle \ldots \rangle$ denotes an average taken over the ensemble of realizations of the scattering medium}. For simplicity, we consider the case of a monodisperse collection of identical scatterers whose positions are independent and identically distributed according to a common density function $p$, which is non-zero only within the slab. Under these assumptions, $C_{ijuv}$ reduces to
\red{\begin{align}\label{eq:char}
	C_{ijuv} \sim \langle \exp(i\mathbf{r}_n\cdot\mathbf{q}) \rangle,
\end{align}
where $\mathbf{q} = \mathbf{k}_i - \mathbf{k}_j - \mathbf{k}_u + \mathbf{k}_v$. Eq.~\eqref{eq:char} can be identified as the characteristic function of the spatial distribution of the scatterers, evaluated at $\mathbf{q}$~\cite{goodman1985statistical}.} We assume that the slab is homogeneous in the transverse coordinates and that \red{particle positions follow the density function $p(\mathbf{r}) = (L_xL_y)^{-1}p_{z}(z)$, where $\mathbf{r}$ is an arbitrary position within the slab and $p_{z}$ is the marginal density function for the $z$ coordinate of a scatterer}. Computing the average in Eq.~(\ref{eq:char}) over the transverse coordinates in the limit $L_x, L_y \to \infty$ yields the delta function $\delta(\mathbf{q}_\perp)$, where $\mathbf{q}_\perp =\mathbf{q}-(\mathbf{q}\cdot\hat{\mathbf{z}})\hat{\mathbf{z}}$ \cite{ByrnesPhD}. This delta function enforces the standard optical memory effect condition $\mathbf{q}_\perp = \mathbf{0}$. Averaging over the $z$ coordinates of the scatterers yields the characteristic function of $p_{z}$, which attains a maximum at $q_z=0$ regardless of the form of $p_z$. Importantly, the memory effect condition $\mathbf{q}_\perp = \mathbf{0}$ does not imply $q_z =0$. Thus, while $\mathbf{q}_\perp = \mathbf{0}$ is a necessary condition for $C_{ijuv} \neq 0$, the magnitude of $C_{ijuv}$ is anisotropic over the set of allowed wavevector combinations, with a maximum attained at $q_z=0.$

\red{While the preceding analysis assumes plane wave illumination, it can be extended to finite width beams. For example, writing the incident field as $E_i(\mathbf r)=a(\mathbf r)\exp(i\mathbf k_i\cdot\mathbf r)$, where $a$ is a spatially varying beam envelope, the average in Eq.~\eqref{eq:char} becomes
\begin{align}
    C_{ijuv} \sim \int p(\mathbf r)
    a(\mathbf r)a'^*(\mathbf r)
    \exp(i\mathbf r \cdot\mathbf q)\,\mathrm{d}^3\mathbf r,
\end{align}
where $a'(\mathbf r)=a(\mathbf R^{-1}\mathbf r)$ is the rotated beam envelope, with $\mathbf R\mathbf k_i=\mathbf k_u$. The covariance is therefore given by the Fourier transform of the scatterer distribution weighted by the beam overlap $a(\mathbf r)a'^*(\mathbf r)$. The finite spatial extent of this overlap broadens the transverse Dirac delta function obtained with plane wave illumination into a correlation peak of finite width, with the original result recovered in the limit of a spatially uniform infinite beam envelope. Although the finite beam width affects the correlation widths, the discussed anisotropy, particularly at $\mathbf q_\perp=\mathbf 0$, is expected to remain qualitatively unchanged for sufficiently broad beams.}

\begin{figure}[t]
	\centering
	\includegraphics[width=\linewidth]{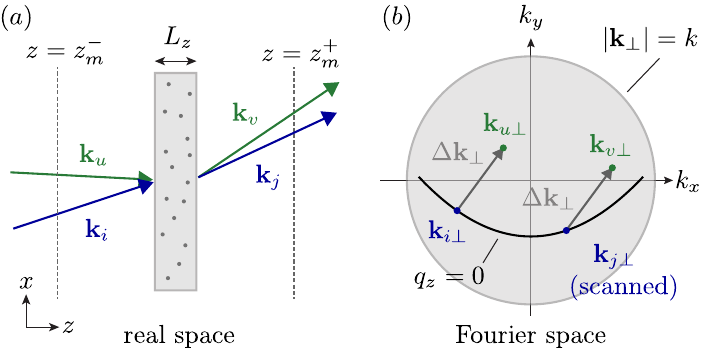}
	\caption{Geometry of the correlated scattering processes. (a) Two scattering processes are compared: $\mathbf{k}_i \to \mathbf{k}_j$ (blue) and $\mathbf{k}_u \to \mathbf{k}_v$ (green), for a slab of thickness $L_z$ containing randomly positioned point scatterers. Scattered fields are evaluated in the measurement planes $z=z_m^{\pm}$, whereby the choice of plane and propagation directions fixes the signs $(s_i,s_j,s_u,s_v)$ in Eq.~(\ref{eq:z-condition}). (b) The same two processes in the transverse Fourier plane. The incident pair is related by the tilt $\Delta\mathbf{k}_\perp = \mathbf{k}_{u\perp}-\mathbf{k}_{i\perp}$, and the memory effect condition $\mathbf{q}_\perp=\mathbf{0}$ requires the scattered pair to be related by the same shift, i.e. $\mathbf{k}_{v\perp}=\mathbf{k}_{j\perp}+\Delta\mathbf{k}_\perp$. In the correlation maps of Figs.~\ref{fig:regular} and \ref{fig:pseudo}, $\mathbf{k}_{i\perp}$ and $\mathbf{k}_{u\perp}$ are held fixed while $\mathbf{k}_{j\perp}$ is scanned over the plane, with $\mathbf{k}_{v\perp}$ following it. Correlations are maximized where $\mathbf{k}_{j\perp}$ additionally lies on the curve $q_z=0$, which necessarily passes through the auto-correlation point $\mathbf{k}_{j\perp}=\mathbf{k}_{i\perp}$.}
	\label{fig:geometry}
\end{figure}
\begin{figure*}[t]
	\centering
	\includegraphics[width=\linewidth]{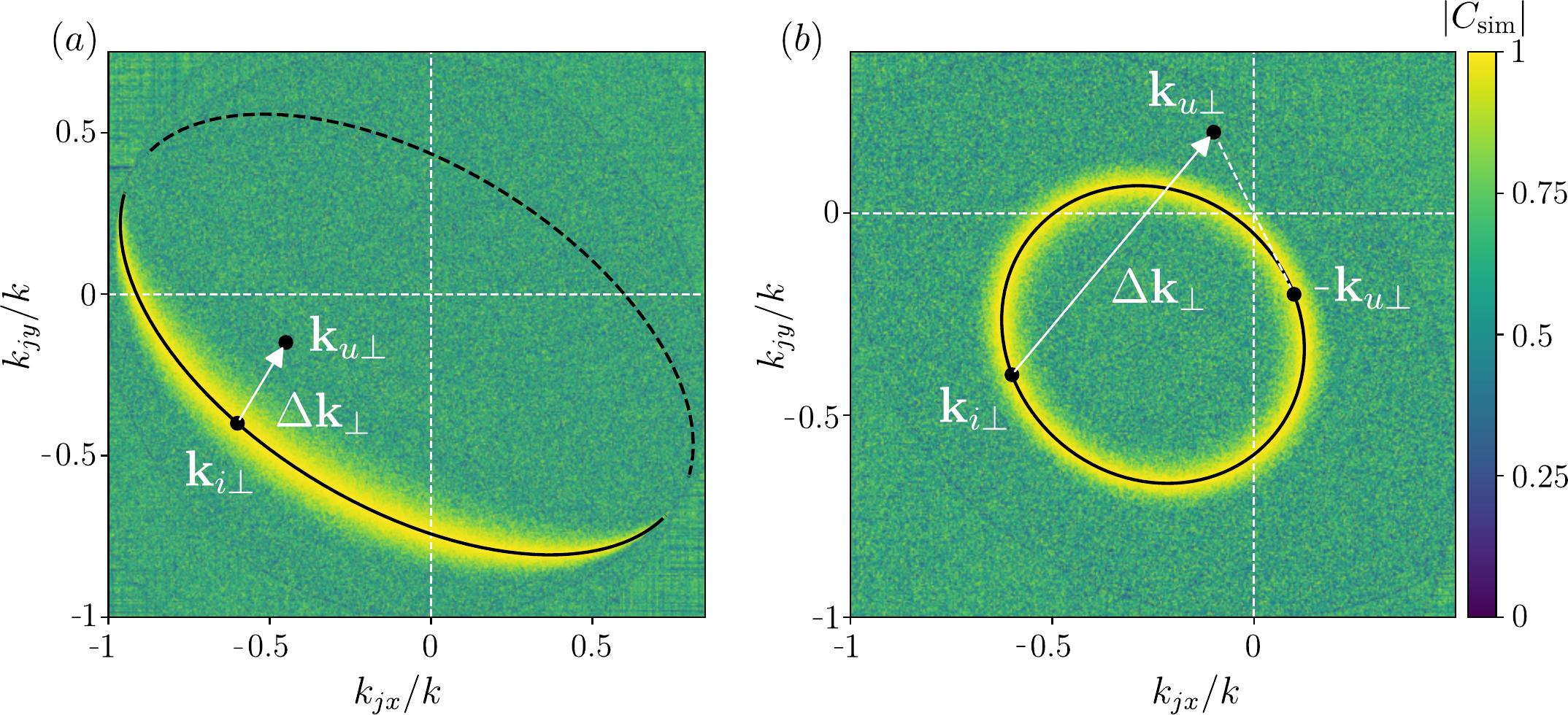}
	\caption{\red{$|C_\mathrm{sim}|$} calculated as a function of $\mathbf{k}_j$ as described in the text. Black circles highlight key wavevectors and black lines show theoretical predictions for the curve $q_z=0$. Correlations are calculated between scattered fields (a) in the same plane $z=z^+_m$ in transmission and (b) between fields evaluated in the planes $z = z_m^{\pm}$, corresponding to two transmitted fields propagating in opposite directions.}
	\label{fig:regular}
\end{figure*}

The set of wavevectors satisfying $q_z=0$ is non-trivial, but can be visualized by considering a simple, hypothetical experiment. Suppose that the slab is illuminated in succession by two plane waves $E_i$ and $E_u$ whose wavevectors $\mathbf{k}_i$ and $\mathbf{k}_u$ are related by a transverse tilt $\Delta\mathbf{k}_\perp = \mathbf{k}_{u\perp} - \mathbf{k}_{i\perp}$. The scattered field in the direction $\mathbf{k}_j$ under illumination by $E_i$ will be correlated with the scattered field in the direction $\mathbf{k}_v$ under illumination by $E_u$ provided that the scattered wavevectors are related by the same transverse tilt, i.e., $\mathbf{k}_{v\perp} - \mathbf{k}_{j\perp} = \Delta\mathbf{k}_\perp$. This geometry, and the roles played by each of the four wavevectors, are summarized in Fig.~\ref{fig:geometry}. Since $|k_z| = \sqrt{k^2-|\mathbf{k}_\perp|^2}$, where $k=2\pi/\lambda$ and $\lambda$ is the wavelength, the additional constraint $q_z=0$ then requires
\begin{align}\label{eq:z-condition}
	\begin{split}
		s_i\sqrt{k^2 - |\mathbf{k}_{i\perp}|^2}& - s_j\sqrt{k^2 - |\mathbf{k}_{j\perp}|^2}\\
		- s_u&\sqrt{k^2 - |\mathbf{k}_{i\perp} + 
			\Delta\mathbf{k}_\perp|^2} \\  &+ s_v\sqrt{k^2 - |\mathbf{k}_{j\perp} + \Delta\mathbf{k}_\perp|^2} =0,
	\end{split}
\end{align}
where, for example, $s_i = \mathrm{sgn}(k_{iz})$, and the signs $s_i, s_j, s_u$ and $s_v$ depend on the nature of the experiment. If, for example, the incident plane waves both propagate in the positive $z$ direction and the scattered fields are measured in reflection, then $(s_i,s_j,s_u,s_v) = (1,-1,1,-1)$. For fixed $\mathbf{k}_{i\perp}$ and $\Delta\mathbf{k}_\perp$, Eq.~(\ref{eq:z-condition}) can be viewed as a non-linear equation for the transverse measurement direction $\mathbf{k}_{j\perp}$. Solutions lie on a curve of maximally-correlated scattering directions in Fourier space. When $\mathbf{k}_j$ does not satisfy Eq.~(\ref{eq:z-condition}), the magnitude of $C_{ijuv}$ is reduced, but does not necessarily vanish. The rate at which $C_{ijuv}$ decreases away from $q_z=0$ is determined by the characteristic function of $p_z$. If, for example, $p_z(z)=L_z^{-1}\mathrm{rect}(z/L_z)$, then \red{$C_{ijuv} \sim \mathrm{sinc}(q_z L_z/2)$}. \red{Thus, the covariance magnitude generally decreases as either $|q_z|$ or $L_z$ increases, where $q_z$ is determined by the geometry of the incident and scattered wavevectors, and $L_z$ is the thickness of the scattering medium. Axial decorrelation can therefore be neglected when $|q_z|L_z \ll 1$.}

\section{Numerical simulations}

To demonstrate the anisotropy of $C_{ijuv}$, we simulate scattering of plane waves by the slab of scatterers described above. We set $\lambda = 500$~nm and define a slab with size $L_x=L_y = 1000\lambda$ and $L_z=10\lambda$ containing $N=1000$ scatterers with positions sampled uniformly over the slab volume. For an incident plane wave $E_i$, we compute the scattered field at an arbitrary point $\mathbf{r}$ away from the slab by summing the first order  scattering from each scatterer according to
\begin{align}\label{eq:cs}
	E_s(\mathbf{r}) =  \sum_{n=1}^{N}\frac{\exp(ik|\mathbf{r}-\mathbf{r}_n|)}{k|\mathbf{r}-\mathbf{r}_n|}E_i(\mathbf{r}_n).
\end{align}
The form of the scattered field in Eq.~(\ref{eq:cs}) is generally appropriate for $\mathbf{r}$ in the far field of each scatterer~\cite{hulst1981light} and implies a single scatterer scattering cross section of $\sigma = 4\pi/k^2$, and thus a mean free path $l=L_xL_yL_z/(\sigma N) \gg L_z$, so that the parameters chosen are consistent with the single scattering regime. We calculate the scattered field in the planes $z^{\pm}_m = \pm(L_z/2 + \delta)$, with $\delta >0$ (set to $5\lambda$ in our simulations) chosen to ensure that $z=z^\pm_m$ is in the far field of the scatterers, over a transverse window centered at the origin and of extent $1000\lambda$ in each transverse direction. The angular spectrum  representation of the scattered field is then found by taking the Fourier transform of the scattered field over the transverse window. The window size was chosen to capture a large number of speckles, ensuring good resolution in the Fourier domain. In addition, full angular coverage in the Fourier domain is achieved by sampling the field over the transverse window at the Nyquist rates $\Delta x = \Delta y =\lambda /2$. Given our parameters, the resulting angular spectra are arrays of size $2000\times 2000$ with $k_x$ and $k_y$ spanning from $-k$ to $k$.

Scattered angular spectrum arrays $\widetilde{E}_j$ and $\widetilde{E}_v$ are first computed for incident plane waves $E_i$ and $E_u$ respectively, with incident transverse wavevectors $\mathbf{k}_{i\perp} = (-0.6, -0.4)^{\mathrm{T}}k$ and $\mathbf{k}_{u\perp} = (-0.45, -0.15)^{\mathrm{T}}k$, corresponding to $\Delta \mathbf{k}_\perp = (0.15, 0.25)^\mathrm{T}k$. This particular wavevector shift corresponds to a pixel shift of  $(\Delta r, \Delta c) = (250, 150)$ rows and columns in our arrays. To facilitate computation of field correlations, we first align the angular spectra by carefully cropping $\widetilde{E}_j$ and $\widetilde{E}_v$. Since $\Delta r, \Delta c > 0$, we remove the top $\Delta r$ rows and final $\Delta c$ columns of $\widetilde{E}_j$, and the bottom $\Delta r$ rows and first $\Delta c$ columns of $\widetilde{E}_v$. The resulting arrays are both of shape $1750\times 1850$ and elements at the same array location are those that are expected to be correlated according to the memory effect condition. To quantify the local correlation between the aligned fields, we define a sliding window of size $t \times t$, which extracts submatrices $W_j$ and $W_v$ of $\widetilde{E}_j$ and $\widetilde{E}_v$ respectively. For each window position, we compute the normalized correlation
\begin{align}\label{eq:C-equation}
	\red{C_\mathrm{sim}} = \frac{\mathrm{tr}(W_j^\dagger W_v)}
	{\sqrt{\mathrm{tr}(W_j^\dagger W_j)
			\mathrm{tr}(W_v^\dagger W_v)}},
\end{align}
where $\mathrm{tr}$ denotes the matrix trace and $\dagger$ is the conjugate transpose operator. \red{We observed that increasing the window size improves statistical averaging, but broadens the apparent width of the correlation features. Conversely, smaller windows provide better spatial resolution, but exhibit larger statistical fluctuations. We found that $t=3$ provides a reasonable compromise between these competing effects, allowing the locations of the correlation peaks to be clearly resolved.}

\red{We emphasize that $C_{\mathrm{sim}}$ is not identical to the theoretical covariance $C_{ijuv}$ defined earlier. In particular, the incident field is excluded, the mean scattered field is not explicitly subtracted, and ensemble averaging is replaced by spatial averaging over a finite window. The mean field contributions are expected to be concentrated within a narrow angular region around the forward scattering direction, with a width determined by the transverse extent of the scattering medium. While these differences can affect the local correlation magnitude, they are not expected to alter the larger scale geometric structures in Fourier space. We therefore use $C_{\mathrm{sim}}$ as a numerical proxy for identifying the predicted anisotropic correlation bands, with spatial averaging allowing these structures to be observed from a single realization of the scattering medium.}

Fig.~\ref{fig:regular}(a) shows \red{$|C_\mathrm{sim}|$} computed for incident plane waves propagating in the positive $z$ direction, i.e. $k_{iz}, k_{uz} > 0$, and $\widetilde{E}_j$ and $\widetilde{E}_v$ measured in the plane $z=z_m^+$, i.e. in transmission. The incident transverse wavevectors $\mathbf{k}_{i\perp}$ and $\mathbf{k}_{u\perp}$ are depicted by black circles and $\Delta\mathbf{k}_\perp$ by a white arrow. A bright band where \red{$|C_\mathrm{sim}| \approx 1$} can be seen to extend across Fourier space, passing through $\mathbf{k}_{i\perp}$ in a direction approximately orthogonal to $\Delta \mathbf{k}_\perp$. The solid black line is the curve $q_z=0$, calculated numerically from Eq.~(\ref{eq:z-condition}) with the choice of signs $(s_i,s_j,s_u,s_v) = (1,1,1,1)$, and can be seen to agree well with the simulated data. 

We note that the curve $q_z=0$ depends on where the scattered field is calculated. If, for example, the scattered fields were instead calculated in the plane $z =z_m^-$, i.e., in reflection, the appropriate set of signs in Eq.~(\ref{eq:z-condition}) would be $(s_i,s_j,s_u,s_v) = (1,-1,1,-1)$. As shown by the dashed black line in Fig.~\ref{fig:regular}(a), the corresponding curve for $q_z = 0$  exhibits opposite curvature to the transmission case.

It is not required that the correlation function be evaluated with two scattered fields in the same plane. In Fig.~\ref{fig:regular}(b), we \red{again show $|C_\mathrm{sim}|$ as in Fig.~\ref{fig:regular}(a), but instead for a case where} $k_{iz} > 0$, $k_{uz} < 0$, and $\widetilde{E}_j$ and $\widetilde{E}_v$ measured in the planes $z=z^{+}_m$ and $z=z^{-}_m$ respectively. In this case, $\widetilde{E}_j$ and $\widetilde{E}_v$ correspond to fields transmitted through the scattering medium in opposite directions. As before, we see a bright band for which \red{$|C_\mathrm{sim}| \approx 1$} passing through $\mathbf{k}_{i\perp}$. In this case, however, the band forms a closed curve. Good agreement is found with the solid black curve computed with Eq.~(\ref{eq:z-condition}) with $(s_i,s_j,s_u,s_v) = (1,1,-1,-1)$. 

When correlating fields that transmit through the slab in opposite directions, we find that $-\mathbf{k}_{u\perp}$ always lies on the curve $q_z=0$. To see why this is the case, note that if $\mathbf{k}_{j\perp} = -\mathbf{k}_{u\perp}$, then $\mathbf{k}_{v\perp} = \mathbf{k}_{j\perp} + \Delta\mathbf{k}_\perp = -\mathbf{k}_{i\perp}$. \red{$|C_\mathrm{sim}|$} thus represents the degree of correlation between the scattering amplitudes associated with scattering between the wavevector pairs $\mathbf{k}_{i\perp} \to \mathbf{k}_{j\perp} = -\mathbf{k}_{u\perp}$ and $\mathbf{k}_{u\perp} \to \mathbf{k}_{v\perp} = -\mathbf{k}_{i\perp}$. These scattering amplitudes are precisely those that are linked by reciprocity, and thus possess a  deterministic relation, regardless of the 
exact form of the scattering medium \cite{Byrnes2021a}. The point $\mathbf{k}_{j\perp} = -\mathbf{k}_{u\perp}$ is in fact a manifestation of the time-reversed memory effect \cite{PhysRevB.41.2635}. Interestingly, the time-reversed peak sits opposite the auto-correlation peak at $\mathbf{k}_{j \perp} = \mathbf{k}_{i\perp}$ on the curve $q_z = 0$.

Finally, we demonstrate that the theory applies equally well to the \red{pseudo-covariance $C_{ijuv}^P = \langle \widetilde{E}_{ji}\widetilde{E}_{vu}\rangle - \langle \widetilde{E}_{ji}\rangle \langle \widetilde{E}_{vu}\rangle$}, which differs from $C_{ijuv}$ by the absence of a complex conjugation on $\widetilde{E}_{vu}$. While $C^P_{ijuv}$ does not exhibit the traditional memory effect, it does exhibit the so-called conjugate memory effect and is non-zero when $\mathbf{q}^P_{\perp}=\mathbf{k}_{i\perp} - \mathbf{k}_{j\perp} + \mathbf{k}_{u\perp} - \mathbf{k}_{v\perp} = \mathbf{0}$~\cite{Kawanishi:99}. This can be derived following the same argument as for the conventional memory effect given above. The only key difference is that the lack of a complex conjugation gives rise to a different set of signs in Eq.~(\ref{eq:char}). To better understand this effect, note that $\mathbf{q}^P_\perp =\mathbf{0}$ implies that $(\mathbf{k}_{i\perp} + \mathbf{k}_{u\perp})/2 =(\mathbf{k}_{j\perp} + \mathbf{k}_{v\perp})/2$. Thus, given a pair of incident wavevectors $\mathbf{k}_{i\perp}$ and $\mathbf{k}_{u\perp}$ with mean value $\bar{\mathbf{k}}_{\perp} = (\mathbf{k}_{i\perp} + \mathbf{k}_{u\perp})/2$, pseudo-correlations exist when $\mathbf{k}_{j\perp}$ and $\mathbf{k}_{v\perp}$ are positioned symmetrically about $\bar{\mathbf{k}}_\perp$. In other words, rather than thinking of $\widetilde{E}_v$ as a tilted version of $\widetilde{E}_j$, with regard to pseudo-correlations, $\widetilde{E}_v$ can be thought of as an inversion of $\widetilde{E}_j$ through the point $\bar{\mathbf{k}}_\perp$. $C^P_{ijuv}$ is also anisotropic, with correlations maximized along the curve $q^P_z=0$, which can be expressed in terms of $\mathbf{k}_{j\perp}$ as
\begin{align}\label{eq:z-pseudo-condition}
	\begin{split}
		s_i\sqrt{k^2 - |\mathbf{k}_{i\perp}|^2}& - s_j\sqrt{k^2 - |\mathbf{k}_{j\perp}|^2}\\
		+ s_u&\sqrt{k^2 - |2\bar{\mathbf{k}}_\perp - 
			\mathbf{k}_{i\perp}|^2} \\  &- s_v\sqrt{k^2 - |2\bar{\mathbf{k}}_\perp - 
			\mathbf{k}_{j\perp}|^2} =0.
	\end{split}
\end{align}

\begin{figure}[t]
	\centering
	\includegraphics[width=\linewidth]{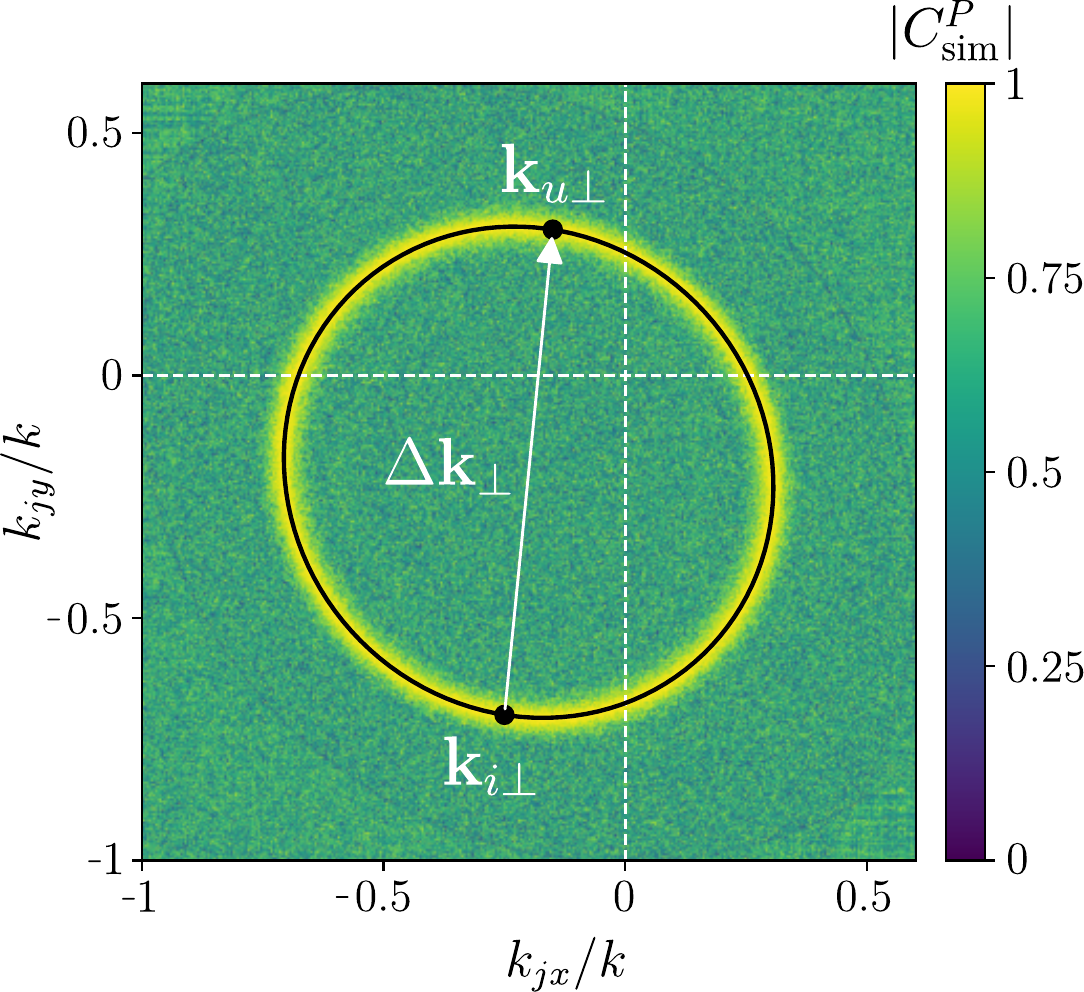}
	\caption{\red{$|C^P_\mathrm{sim}|$} calculated as a function of $\mathbf{k}_j$ as described in the text. Black circles highlight key wavevectors and the black line shows the theoretical prediction for the curve $q^P_z=0$. Correlations are calculated between scattered fields in the same plane $z=z^+_m$ in transmission.}
	\label{fig:pseudo}
\end{figure}

Setting $\mathbf{k}_{i\perp} = (-0.25,-0.7)^{\mathrm{T}}k$ and $\mathbf{k}_{u\perp} = (-0.15,0.3)^{\mathrm{T}}k$, corresponding to $\bar{ \mathbf{k}}_\perp = (-0.2,-0.2)^\mathrm{T}k$, we compute \red{$C_\mathrm{sim}^P$} using Eq.~(\ref{eq:C-equation}), but replacing the conjugate transpose with a matrix transpose. In order to align $\widetilde{E}_j$ and $\widetilde{E}_v$, we map each element in $\widetilde{E}_v$ to the location related by an inversion through the array position corresponding to $\bar{\mathbf{k}}_\perp$. The values of array elements whose inverted positions would lie outside of the array are set to 0. 

Fig.~\ref{fig:pseudo} shows \red{$|C_\mathrm{sim}^P|$} in the case $k_{iz}, k_{uz} > 0$ and $\widetilde{E}_j$ and $\widetilde{E}_v$ calculated in the plane $z=z^+_m$, i.e. in transmission. The solid line, computed now from Eq.~(\ref{eq:z-pseudo-condition}) with signs $(s_i, s_j, s_u, s_v) = (1,1,1,1)$, agrees well with the simulated data. We note that while this choice of signs is identical to that used in Fig.~\ref{fig:regular}(a), there are additional sign changes inherent to Eq.~(\ref{eq:z-pseudo-condition}) compared to Eq.~(\ref{eq:z-condition}) that result in the curve's being closed, rather than open as before. We also note that $\mathbf{k}_{u\perp}$ now lies on the curve, which highlights a symmetry between the scattering amplitudes associated with scattering between the wavevector pairs $\mathbf{k}_{i\perp} \to \mathbf{k}_{u\perp}$ and $\mathbf{k}_{u\perp} \to \mathbf{k}_{i\perp}$. This is not a consequence of reciprocity, but is instead a symmetry unique to single scattering~\cite{hulst1981light}.

\section{Discussion}

The anisotropy described above follows from the geometric observation that the transverse and axial components of a propagating wavevector are not independent, but are linked by the dispersion relation $|k_z| = \sqrt{k^2 - |\mathbf{k}_\perp|^2}$. The condition $q_z=0$ can therefore not be satisfied independently of the memory effect condition $\mathbf{q}_\perp = \mathbf{0}$. Once $\mathbf{q}_\perp = \mathbf{0}$ is imposed, $q_z$ depends only on the transverse wavevectors, so that $q_z=0$ defines a one-dimensional curve within the two-dimensional set of accessible measurement directions $\mathbf{k}_{j\perp}$. Importantly, as seen in Figs.~\ref{fig:regular} and \ref{fig:pseudo}, the shape of this curve is not universal, but is determined by the measurement geometry through the signs $(s_i,s_j,s_u,s_v)$. The $q_z=0$ curve is, for example, open in transmission, open with opposite curvature in reflection, and closed when the two scattered fields propagate in opposite directions. Critically, however, in all cases the memory is strongest along a single curve of correlated scattering directions and is not isotropic in Fourier space. This picture also connects several correlation phenomena that are usually treated separately. Specifically, the conventional angular memory effect, the time-reversed memory effect and the conjugate memory effect were each found to correspond to a point (or points) on appropriately chosen $q_z=0$ curves. 

Although our simulation results were presented assuming a uniformly axially disordered slab for simplicity, the underlying physics is notably more general. In particular, our derivations were expressed in terms of the characteristic function of an arbitrary marginal density $p_z$ describing the axial disorder. The specific form of the axial density $p_z$  affects only the width and shape of the high correlation band, but not the location of its maximum. In principle, this dependence could therefore be used to recover information about the medium thickness and the axial distribution of scatterers from measurements of the correlation function. 

Single scattering has been assumed throughout this work, thereby limiting the domain of validity of our findings. For higher order scattering trajectories, the phase of a partial field is no longer set by the position of a single scatterer, and $\mathbf{q}$ is replaced by a sum of contributions from each segment of the path. \red{While the transverse memory effect is known to persist in the multiple scattering regime, the evolution of the residual covariance function with increasing scattering order is less clear. We conjecture that the predicted anisotropy may persist in weakly scattering media beyond the single scattering regime, but may become progressively less pronounced as the scattering order increases.} A quantitative treatment of this regime however lies beyond the scope of this work.

Practically, the observed anisotropy has consequences for measurements that exploit the memory effect over a wide angular range. Since correlations decay away from the $q_z=0$ condition, a measurement that scans $\mathbf{k}_{j\perp}$ at a fixed tilt will record a loss of correlation that does not arise from the transverse memory effect condition. For a small numerical aperture setup, the accessible range of $\mathbf{k}_{j\perp}$ is small and the curve covers most of the measurement region, so that this distinction is unimportant. Indeed, the reported anisotropy has likely not been observed to date, since  most memory effect related experiments operate close to the paraxial regime about normal incidence, for which $q_z \approx 0$ is automatically satisfied. For a large numerical aperture experiment, or for large tilts, however, the decay is significant and could easily be misinterpreted as a reduced memory effect range. We expect this to be of particular relevance to wavefront shaping over large solid angles.

Finally, we note that although $C_{ijuv}$ and $C^P_{ijuv}$ are field \red{covariances}, they can be accessed through intensity measurements, presenting a potential route to experimental observation of the predicted anisotropy. Note that a non-zero $C^P_{ijuv}$ implies that the scattered field is not circularly symmetric in the complex plane. For jointly Gaussian, non-circular fields \red{with vanishing mean}, the moment theorem gives
\red{\begin{align}\label{eq:siegert}
	\langle I_{ji} I_{vu}\rangle = \langle I_{ji}\rangle\langle I_{vu}\rangle + |C_{ijuv}|^2 + |C^P_{ijuv}|^2,
\end{align}}which generalizes the Siegert relation by retaining the pseudo-\red{covariance} term. \red{Intensity correlations measured in the Fourier domain should therefore exhibit both structures described above, with each contribution appearing when its corresponding transverse wavevector phase matching condition is satisfied.} Observing this structure would however require the correlation to be resolved as a function of $\mathbf{k}_{j\perp}$ over a wide angular range, whilst also using a sample thin enough that the high correlation band can be resolved.

\section{Conclusions}

To conclude, we have shown that memory effect correlations are anisotropic in the Fourier domain. In particular, correlations are maximized when the axial components of the relevant set of wavevectors satisfy a similar equation to that normally satisfied by the transverse components. This phenomenon can be explained straightforwardly in the single scattering regime. Numerical simulations support our results and show strong agreement with theoretical predictions. Overall, this work provides a deeper understanding of speckle correlations in random scattering media and reveals that speckle patterns encode additional information about the underlying scatterer distribution than has previously been reported.

\section*{CRediT authorship contribution statement}
\textbf{Niall Byrnes:} Conceptualization, Methodology, Software, Formal analysis, Writing - original draft. \textbf{Matthew R. Foreman:} Conceptualization, Software, Supervision, Writing - review \& editing, Funding acquisition.

\section*{Declaration of competing interest}
The authors declare that they have no known competing financial interests or personal relationships that could have appeared to influence the work reported in this paper.

\section*{Data availability}
The data that support the findings of this study are available from the corresponding author upon reasonable request.

\section*{Acknowledgements}
	N.B. and M.R.F. were supported by the Nanyang Technological University Grant SUG:022824-00001. M.R.F. acknowledges additional funding from the Institute for Digital Molecular Analytics and Science (IDMxS) under the Singapore Ministry of Education Research Centres of Excellence scheme (EDUN C-33-18-279-V12).


\begin{thebibliography}{10}
\expandafter\ifx\csname url\endcsname\relax
  \def\url#1{\texttt{#1}}\fi
\expandafter\ifx\csname urlprefix\endcsname\relax\def\urlprefix{URL }\fi
\expandafter\ifx\csname href\endcsname\relax
  \def\href#1#2{#2} \def\path#1{#1}\fi

\bibitem{PhysRevLett.61.834}
S.~Feng, C.~Kane, P.~A. Lee, A.~D. Stone,
  \href{https://link.aps.org/doi/10.1103/PhysRevLett.61.834}{Correlations and
  fluctuations of coherent wave transmission through disordered media}, Phys.
  Rev. Lett. 61 (1988) 834--837.
\newblock \href {https://doi.org/10.1103/PhysRevLett.61.834}
  {\path{doi:10.1103/PhysRevLett.61.834}}.
\newline\urlprefix\url{https://link.aps.org/doi/10.1103/PhysRevLett.61.834}

\bibitem{Liu:19}
H.~Liu, Z.~Liu, M.~Chen, S.~Han, L.~V. Wang,
  \href{https://opg.optica.org/prj/abstract.cfm?URI=prj-7-11-1323}{Physical
  picture of the optical memory effect}, Photon. Res. 7~(11) (2019) 1323--1330.
\newblock \href {https://doi.org/10.1364/PRJ.7.001323}
  {\path{doi:10.1364/PRJ.7.001323}}.
\newline\urlprefix\url{https://opg.optica.org/prj/abstract.cfm?URI=prj-7-11-1323}

\bibitem{PhysRevB.41.2635}
R.~Berkovits, M.~Kaveh,
  \href{https://link.aps.org/doi/10.1103/PhysRevB.41.2635}{Time-reversed memory
  effects}, Phys. Rev. B 41 (1990) 2635--2638.
\newblock \href {https://doi.org/10.1103/PhysRevB.41.2635}
  {\path{doi:10.1103/PhysRevB.41.2635}}.
\newline\urlprefix\url{https://link.aps.org/doi/10.1103/PhysRevB.41.2635}

\bibitem{Berkovits_1990}
R.~Berkovits, M.~Kaveh, \href{https://doi.org/10.1209/0295-5075/13/2/001}{The
  vector memory effect for waves}, EPL 13~(2) (1990) 97.
\newblock \href {https://doi.org/10.1209/0295-5075/13/2/001}
  {\path{doi:10.1209/0295-5075/13/2/001}}.
\newline\urlprefix\url{https://doi.org/10.1209/0295-5075/13/2/001}

\bibitem{PhysRevLett.123.203901}
H.~Y{\i}lmaz, C.~W. Hsu, A.~Goetschy, S.~Bittner, S.~Rotter, A.~Yamilov,
  H.~Cao,
  \href{https://link.aps.org/doi/10.1103/PhysRevLett.123.203901}{Angular memory
  effect of transmission eigenchannels}, Phys. Rev. Lett. 123 (2019) 203901.
\newblock \href {https://doi.org/10.1103/PhysRevLett.123.203901}
  {\path{doi:10.1103/PhysRevLett.123.203901}}.
\newline\urlprefix\url{https://link.aps.org/doi/10.1103/PhysRevLett.123.203901}

\bibitem{Katz2014}
O.~Katz, P.~Heidmann, M.~Fink, S.~Gigan,
  \href{http://dx.doi.org/10.1038/nphoton.2014.189}{Non-invasive single-shot
  imaging through scattering layers and around corners via speckle
  correlations}, Nat. Photon. 8~(10) (2014) 784–790.
\newblock \href {https://doi.org/10.1038/nphoton.2014.189}
  {\path{doi:10.1038/nphoton.2014.189}}.
\newline\urlprefix\url{http://dx.doi.org/10.1038/nphoton.2014.189}

\bibitem{10.1117/1.AP.8.3.036009}
X.~Luo, J.~Liu, X.~Wang, J.~Li, M.~Xiang, D.~Wang, J.~Duan, T.~Zhang, X.~Dong,
  Z.~Geng, F.~Liu, \href{https://doi.org/10.1117/1.AP.8.3.036009}{{Single-shot
  scattering imaging beyond the optical memory effect via speckle-sparsity
  decoupling}}, Advanced Photonics 8~(3) (2026) 036009.
\newblock \href {https://doi.org/10.1117/1.AP.8.3.036009}
  {\path{doi:10.1117/1.AP.8.3.036009}}.
\newline\urlprefix\url{https://doi.org/10.1117/1.AP.8.3.036009}

\bibitem{PhysRevX.11.031010}
H.~Y{\i}lmaz, M.~K\"uhmayer, C.~W. Hsu, S.~Rotter, H.~Cao,
  \href{https://link.aps.org/doi/10.1103/PhysRevX.11.031010}{Customizing the
  angular memory effect for scattering media}, Phys. Rev. X 11 (2021) 031010.
\newblock \href {https://doi.org/10.1103/PhysRevX.11.031010}
  {\path{doi:10.1103/PhysRevX.11.031010}}.
\newline\urlprefix\url{https://link.aps.org/doi/10.1103/PhysRevX.11.031010}

\bibitem{Osnabrugge:17}
G.~Osnabrugge, R.~Horstmeyer, I.~N. Papadopoulos, B.~Judkewitz, I.~M.
  Vellekoop,
  \href{https://opg.optica.org/optica/abstract.cfm?URI=optica-4-8-886}{Generalized
  optical memory effect}, Optica 4~(8) (2017) 886--892.
\newblock \href {https://doi.org/10.1364/OPTICA.4.000886}
  {\path{doi:10.1364/OPTICA.4.000886}}.
\newline\urlprefix\url{https://opg.optica.org/optica/abstract.cfm?URI=optica-4-8-886}

\bibitem{Zhang:21}
R.~Zhang, J.~Du, Y.~He, D.~Yuan, J.~Luo, D.~Wu, B.~Ye, Z.-C. Luo, Y.~Shen,
  \href{https://opg.optica.org/oe/abstract.cfm?URI=oe-29-17-26944}{Characterization
  of the spectral memory effect of scattering media}, Opt. Express 29~(17)
  (2021) 26944--26954.
\newblock \href {https://doi.org/10.1364/OE.434331}
  {\path{doi:10.1364/OE.434331}}.
\newline\urlprefix\url{https://opg.optica.org/oe/abstract.cfm?URI=oe-29-17-26944}

\bibitem{Zhu:20}
L.~Zhu, J.~B. de~Monvel, P.~Berto, S.~Brasselet, S.~Gigan, M.~Guillon,
  \href{https://opg.optica.org/optica/abstract.cfm?URI=optica-7-4-338}{Chromato-axial
  memory effect through a forward-scattering slab}, Optica 7~(4) (2020)
  338--345.
\newblock \href {https://doi.org/10.1364/OPTICA.382209}
  {\path{doi:10.1364/OPTICA.382209}}.
\newline\urlprefix\url{https://opg.optica.org/optica/abstract.cfm?URI=optica-7-4-338}

\bibitem{Amitonova:15}
L.~V. Amitonova, A.~P. Mosk, P.~W.~H. Pinkse,
  \href{https://opg.optica.org/oe/abstract.cfm?URI=oe-23-16-20569}{Rotational
  memory effect of a multimode fiber}, Opt. Express 23~(16) (2015)
  20569--20575.
\newblock \href {https://doi.org/10.1364/OE.23.020569}
  {\path{doi:10.1364/OE.23.020569}}.
\newline\urlprefix\url{https://opg.optica.org/oe/abstract.cfm?URI=oe-23-16-20569}

\bibitem{10.1063/5.0303181}
U.~Gokay, R.~J. Kilpatrick, S.~A.~R. Horsley, D.~B. Phillips, J.~Bertolotti,
  \href{https://doi.org/10.1063/5.0303181}{Explaining and exploiting the radial
  memory effect in multimode optical fibers}, APL Photonics 11~(2) (2026)
  026107.
\newblock \href
  {http://arxiv.org/abs/https://pubs.aip.org/aip/app/article-pdf/doi/10.1063/5.0303181/20906752/026107_1_5.0303181.pdf}
  {\path{arXiv:https://pubs.aip.org/aip/app/article-pdf/doi/10.1063/5.0303181/20906752/026107_1_5.0303181.pdf}},
  \href {https://doi.org/10.1063/5.0303181} {\path{doi:10.1063/5.0303181}}.
\newline\urlprefix\url{https://doi.org/10.1063/5.0303181}

\bibitem{Byrnes02012026}
N.~Byrnes, M.~R. Foreman,
  \href{https://doi.org/10.1080/17455030.2022.2153305}{Random matrix theory of
  polarized light scattering in disordered media}, Waves Random Complex Media
  36~(1) (2026) 249--277.
\newblock \href {https://doi.org/10.1080/17455030.2022.2153305}
  {\path{doi:10.1080/17455030.2022.2153305}}.
\newline\urlprefix\url{https://doi.org/10.1080/17455030.2022.2153305}

\bibitem{ByrnesPhD}
N.~F. Byrnes, Random matrix modelling of polarised light scattering in
  disordered media, Phd, Imperial College London (2023).

\bibitem{ByrnesCPC}
N.~Byrnes, S.~Dutta, M.~R. Foreman,
  \href{https://www.sciencedirect.com/science/article/pii/S0010465526002961}{Extended
  scattering channels for random matrix simulations of polarized light
  transport}, Comp. Phys. Commun. 328 (2026) 110314.
\newblock \href {https://doi.org/https://doi.org/10.1016/j.cpc.2026.110314}
  {\path{doi:https://doi.org/10.1016/j.cpc.2026.110314}}.
\newline\urlprefix\url{https://www.sciencedirect.com/science/article/pii/S0010465526002961}

\bibitem{Mandel_Wolf_1995}
L.~Mandel, E.~Wolf, Optical Coherence and Quantum Optics, Cambridge University
  Press, 1995.

\bibitem{goodman1985statistical}
J.~Goodman,
  \href{https://books.google.com.sg/books?id=2VTwAAAAMAAJ}{Statistical Optics},
  Wiley, 1985.
\newline\urlprefix\url{https://books.google.com.sg/books?id=2VTwAAAAMAAJ}

\bibitem{hulst1981light}
H.~Hulst, H.~van~de Hulst, Light Scattering by Small Particles, Dover
  Publications, 1981.

\bibitem{Byrnes2021a}
N.~Byrnes, M.~R. Foreman, Symmetry constraints for vector scattering and
  transfer matrices containing evanescent components: energy conservation,
  reciprocity and time reversal, Phys. Rev. Research 3 (2021) 013129.
\newblock \href {https://doi.org/10.1103/PhysRevResearch.3.013129}
  {\path{doi:10.1103/PhysRevResearch.3.013129}}.

\bibitem{Kawanishi:99}
T.~Kawanishi, Z.~L. Wang, M.~Izutsu, H.~Ogura,
  \href{https://opg.optica.org/josaa/abstract.cfm?URI=josaa-16-6-1342}{Conjugate
  memory effect of random scattered waves}, J. Opt. Soc. Am. A 16~(6) (1999)
  1342--1349.
\newblock \href {https://doi.org/10.1364/JOSAA.16.001342}
  {\path{doi:10.1364/JOSAA.16.001342}}.
\newline\urlprefix\url{https://opg.optica.org/josaa/abstract.cfm?URI=josaa-16-6-1342}

\end{thebibliography}
\end{document}